# Simulation of Observed Magnetic Holes in the Magnetosheath

**Narges Ahmadi[1], Kai Germaschewski[2], and Joachim Raeder[2]**

[1]Laboratory for Atmospheric and Space Physics, University of Colorado, Boulder, Colorado, USA.

[2]Department of Physics and Space Science Center, University of New Hampshire, Durham, New Hampshire, USA.

Corresponding author: Narges Ahmadi (Narges.Ahmadi@colorado.edu)

**Key Points:**

- Nonlinear evolution of mirror instability leads to magnetic holes in PIC expanding box simulations.
- Plasma is mirror unstable in magnetic holes and it is mirror stable in magnetic peaks.
- In expanding box simulation, plasma follows the marginal stability path of proton cyclotron instability.

**Abstract**

Magnetic holes have been frequently observed in the Earth's magnetosheath and it is believed that these structures are the result of nonlinear evolution of mirror instability. Mirror mode fluctuations mostly appear as magnetic holes in regions where plasma is marginally mirror stable with respect to the linear instability. We present an expanding box particle-in-cell simulation to mimic the magnetosheath plasma and produce the mirror mode magnetic holes. We show that magnetic peaks are dominant when plasma is mirror unstable and mirror fluctuations evolve to deep magnetic holes when plasma is marginally mirror stable. Although, the averaged plasma parameters in the simulation are marginally close to mirror instability threshold, the plasma in the magnetic holes is highly unstable to mirror instability and mirror stable in the magnetic peaks.

**1 Introduction**

The motivation for this study originates from frequent observations of magnetic holes in the planetary magnetosheaths [*Joy et al.*, 2006; *Soucek et al*, 2008; *Génot et al*, 2009; …]. Magnetic holes are sudden drops in the background magnetic field with amplitudes as large as the ambient magnetic field. Statistical and theoretical studies show that the magnetic holes are the result of nonlinear evolution of mirror instability [*Kivelson and Southwood*, 1996; *Soucek et al*, 2008; *Génot et al*, 2009; …]. Mirror instability is generated when there is a proton temperature anisotropy with $T_{p\perp} > T_{p||}$, where the perpendicular and parallel temperatures are relative to the background magnetic field. Mirror instability has zero frequency ($\omega = 0$) in the plasma frame in a homogeneous plasma and its wave vector is oblique to the background magnetic field. The proton temperature anisotropy ($T_{p\perp} > T_{p||}$) also leads to the generation of proton cyclotron instability. In an electron-proton plasma according to linear dispersion theory, the proton cyclotron instability has a lower threshold and larger maximum growth rate than the mirror instability under many space plasma conditions. *Price et al.* [1986] have shown that the presence of small density of heavy ions could reduce the linear growth rate of the proton cyclotron instability, while leaving the mirror instability unchanged. Also, the electron temperature anisotropy ($T_{e\perp} > T_{e||}$) enhances the mirror instability growth rate but does not affect the proton cyclotron instability growth rate significantly [*Gary*, 1992; *Remya et al.*, 2013; *Ahmadi et al.*, 2016a; *Ahmadi et al.*, 2016b]. *Ahmadi et al.* [2016a] have shown that the presence of electron temperature anisotropy ($T_{e\perp} > T_{e||}$) may generate the electron whistler instability which quickly consumes most of the electron free energy before mirror instability can grow.

Satellite observations consistently find mirror mode structures in planetary magnetosheath and the solar wind [*Kaufmann et al.*, 1970; *Tsurutani et al.*, 1982; *Winterhalter et al.*, 1995; *Erdos and Balogh*, 1996; *Bavassano-Cattaneo*, 1998; *Joy et al.*, 2006; *Soucek et al*, 2008; *Balikhin et al.*, 2009; *Dimmock et al.*, 2015; *Osmane et al.*, 2015]. These nonlinear mirror mode structures are typically observed as trains of quasi-periodic magnetic peaks and opposite structures called magnetic holes. In contrast, numerical simulation studies have shown that the nonlinear evolution of mirror instability leads to magnetic peaks [*Califano et al.*, 2008; *Porazik and Johnson*, 2013a]. *Porazik and Johnson* [2013a, 2013b] used gyrokinetic approach to simulate the nonlinear development of the mirror instability and they showed the development of peaked saturated structures. Nonlinear evolution of mirror instability is a topic of active investigation since the observations of magnetic holes have not been fully understood. *Kivelson and Southwood* [1996] discuss that nonlinear evolution of mirror instability leads to magnetic holes because the saturation is provided by the cooling of trapped ion population. *Kuznetsov et al.* [2007] use a reductive perturbative expansion of Vlasov-Maxwell equations and show that magnetic holes are a solution to the equations in a mirror stable plasma but magnetic peaks cannot survive in such conditions. This is called bi-stability theory. Also, their model leads to magnetic peaks for nonlinear saturation of mirror instability. *Califano et al.* [2008] used high resolution hybrid numerical simulations of the Vlasov-Maxwell equations to study the nonlinear mirror mode and they concluded that direct nonlinear saturation of mirror instability leads to magnetic peaks. Therefore, the question is under what conditions the mirror instability evolves to magnetic holes in its nonlinear stage of evolution. Several statistical studies have shown that the shape of mirror structures is related to local plasma parameters [*Joy et al.*, 2006; *Soucek et al.*, 2008]. Specifically, low $\beta_{p||}$ (ratio of plasma to magnetic pressure) conditions are associated with observations of magnetic holes while magnetic peaks are usually observed in higher $\beta_{p||}$ plasma. Satellite observations have shown that the mirror and proton cyclotron instability regulate the plasma in the magnetosheath and put an upper limit on the temperature anisotropy and therefore these instabilities contribute significantly to magnetosheath dynamics. The theoretical mirror instability threshold in a plasma with warm anisotropic protons and electrons is given by [*Pantellini and Schwartz*, 1995; *Pokhotelov et al.*, 2000]:

$$R_m = \beta_{p\perp}\left(\frac{T_{p\perp}}{T_{p\|}} - 1\right) + \beta_{e\perp}\left(\frac{T_{e\perp}}{T_{e\|}} - 1\right) - \frac{\beta_{e\|}}{2}\frac{(T_{p\perp}/T_{p\|} - T_{e\perp}/T_{e\|})}{1 + T_{e\|}/T_{p\|}} - 1 \qquad (1)$$

If $R_m \geq 0$, plasma is unstable to mirror instability and if $R_m < 0$, plasma is stable to mirror instability. Magnetic fluctuations mostly appear as magnetic holes in regions where plasma is marginally stable with respect to the linear instability. *Hellinger et al.* [2003] and *Trávníček et al.* [2007] studied the effect of compression and expansion on the mirror and proton cyclotron instabilities using hybrid simulations and they showed that in their simulations, mirror and proton cyclotron instabilities keep the plasma in a marginally stable state.

The Voyager spacecraft observed mirror structures on the dayside of Saturn as reported by *Bavassano-Cattaneo et al.* [1998]. The evolution of mirror structures from quasi-perpendicular bow shock to the magnetopause shows that mirror structures evolve from quasi-sinusoidal waves downstream of the shock layer to magnetic holes close to the magnetopause. *Soucek et al.* [2008] studied the mirror structures in the Earth's magnetosheath using Cluster spacecraft. They report that magnetic peaks are observed in a mirror unstable plasma while magnetic holes are dominant where the plasma is mirror stable or marginally mirror stable. Using multi-spacecraft analysis, they observe an abrupt transition of mirror structures from peaks to holes at an approximate distance of 2 Earth radii from the magnetopause and they interpret this effect as a consequence of plasma expansion in the vicinity of the magnetopause where plasma conditions are locally changing toward a more mirror stable state. Based on these observations, different models have been proposed to explain how mirror structures evolve from the bow shock to the magnetopause [*Génot et al*, 2009; …]. Several computational and theoretical works have studied the evolution of mirror structures [].

We developed an expanding box particle-in-cell (PIC) simulation in order to resemble the convection of the mirror structures in the magnetosheath. As the plasma expands we can control how to change the plasma parameters according to the direction of the background magnetic field and direction of expansion. This allows us to increase or decrease the temperature anisotropy and $\beta$ in the plasma. Therefore, we can investigate how the mirror structures will evolve in particular their transition from magnetic peaks to magnetic holes as plasma becomes marginally mirror stable.

This paper is organized as follows. Section 2 presents the equations for the PIC method in an expanding box. Section 3 discusses the nonlinear evolution of mirror instability in expanding box simulations and shows that plasma follows the marginal stability path for proton cyclotron instability. Section 4 presents the result of the simulation and shows that mirror instability evolves to magnetic holes in a marginally mirror stable plasma. In section 5, the main results are summarized in the conclusion.

## 2 PIC Method in an Expanding Box

We study the evolution of expansion-driven mirror instability by means of fully-kinetic PIC simulations. We have modified the three-dimensional Plasma Simulation Code (PSC) [*Germaschewski et al.,* 2016] to account for the effect of an overall expansion or compression of the system. We implemented the expanding (compressing) box technique introduced by *Sironi and Narayan* [2015] into PSC. So far, an expanding (compressing) computational domain has been employed to study the behavior of temperature anisotropy driven instabilities in hybrid models with kinetic ions and fluid electrons [*Hellinger et al.*, 2003]. *Sironi and Narayan* [2015] and *Sironi* [2015] have used PIC compressing box to study the electron heating by the proton cyclotron instability in collisonless accretion flows such as black hole in the Milky Way. In a PIC expanding or compressing box method, we solve the Maxwell's and momentum equations in a fluid co-moving frame. The fluid co-moving frame is related to the laboratory frame by a Lorentz transformation, with velocity $\boldsymbol{U}$. There are two sets of spatial coordinates, primed and unprimed coordinate systems, in the fluid co-moving frame. The particle location in the laboratory frame (labeled with subscript $L$) is related to its position in the primed coordinate system by $\boldsymbol{x}_L = \boldsymbol{L}\boldsymbol{x}'$ where compression or expansion are described by a diagonal matrix

$$\boldsymbol{L} = \frac{\partial \boldsymbol{x}}{\partial \boldsymbol{x}'} = \begin{pmatrix} a_x & 0 & 0 \\ 0 & a_y & 0 \\ 0 & 0 & a_z \end{pmatrix}, \quad l = \det(\boldsymbol{L}) \tag{2}$$

where $l$ is the determinant. In general, $a_x$, $a_y$ and $a_z$ are functions of time, but not of the spatial coordinates. Since both primed and unprimed coordinate systems exist in the co-moving frame, they have the same time coordinate $t = t'$ and $dt = dt'$. The relation between the primed and unprimed coordinate is such that $d\boldsymbol{x} = \boldsymbol{L}d\boldsymbol{x}'$ and $\boldsymbol{U} = \dot{\boldsymbol{L}}\boldsymbol{x}' = \dot{\boldsymbol{L}}\boldsymbol{L}^{-1}\boldsymbol{x}_L$. It is reasonable to only consider non-relativistic limit ($|\boldsymbol{U}|/c \ll 1$) since the compression and expansion velocities in the magnetosheath are non-relativistic. In the magnetosphere concept, the velocity $\boldsymbol{U}$ is the expansion or compression velocity of the magnetosheath which depends on solar wind pressure. For example, for an expansion parallel to the background magnetic field aligned in the $z$ direction,

$$a_x = 1, \ a_y = 1, a_z = 1 + q_z t$$

where $1/q_z$ is the expansion characteristic time. Therefore, in our setup, $U_x = U_y = 0$, whereas $U_z = q_z z'$. The typical scale of our simulations is proton Larmor radius $\rho_p = v_{th,p}/\omega_p$ and this gives us $U_z/c \sim q_z \rho_p/c \sim (q_z/\omega_p)(v_{th,p}/c)$, where $v_{th,p}$ is the proton thermal velocity and $\omega_p$ refers to the proton plasma frequency. In the magnetosheath regime, we expect non-relativistic protons $(v_{th,p}/c \ll 1)$ and slow expansion $(q_z/\omega_p \ll 1)$, so our assumption $U/c \ll 1$ is easily satisfied.

The Ampere's law and Faraday law in the primed coordinate are,

$$\nabla' \times (l^{-1}\mathbf{L}^2 \mathbf{E}') = -\frac{1}{c}\frac{\partial \mathbf{B}'}{\partial t'}$$

$$\nabla' \times (l^{-1}\mathbf{L}^2 \mathbf{B}') = \frac{1}{c}\frac{\partial \mathbf{E}'}{\partial t'} + \frac{4\pi}{c} l\mathbf{J}'$$

We implement the momentum equation in the unprimed coordinate system and the particle push in the primed coordinate system in the algorithm. The evolution of the particle orbits is solved with these set of equations,

$$\frac{d\mathbf{p}}{dt} = -\dot{\mathbf{L}}\mathbf{L}^{-1}\mathbf{p} + q\left(\mathbf{E} + \frac{\mathbf{v}}{c}\times\mathbf{B}\right)$$

$$\frac{d\mathbf{x}'}{dt'} = \mathbf{v}'$$

The current density $\mathbf{J}'$ and charge density $\rho'$ in the primed coordinate system is given by

$$\mathbf{J}' = L^{-1}\mathbf{J} = l^{-1}\sum_\alpha q_\alpha \mathbf{v}_\alpha S[\mathbf{x}' - \mathbf{x}'_\alpha(t')]$$

$$\rho' = l^{-1}\sum_\alpha q_\alpha S[\mathbf{x}' - \mathbf{x}'_\alpha(t')]$$

These are the set of equations that we solve in the primed and unprimed coordinates. We measure the physical quantities in the unprimed coordinate system since it has a basis of unit vectors. The post processing of the quantities transform them into laboratory frame which is used in this paper.

## 3 Nonlinear Evolution in Expanding Box Simulation

In this section, we use the implemented expanding box method into our PIC code to study the evolution of expansion-driven mirror instability. As we mentioned earlier, we want to mimic the plasma expansion in the magnetosheath. The expanding box simulation models the evolution of magnetosheath plasma which expands under the effect of the global magnetosheath flow and changes in solar wind dynamic pressure. This is a self-consistent way to study the properties of the waves driven by temperature anisotropy in the magnetosheath plasma. We perform 2-dimensional expanding box simulation with an expansion along the background magnetic field in the z direction (parallel expansion). According to CGL (Chew, Goldberger and Low) condition [*Chew et al.,* 1956], for an expansion along the background magnetic field, the conservation of first and second adiabatic invariants leads to,

$$\frac{T_\perp}{T_\parallel} \propto L^2, \beta_\perp \propto \frac{1}{L}, \beta_\parallel \propto \frac{1}{L^3}$$

Therefore, as $L$ increases the temperature anisotropy increases while plasma $\beta$ decreases. *Génot et al.* [2009] have used hybrid expanding box model to study the mirror instability evolution. In their model, the electrons are treated as a fluid and isothermal. In our simulation, the electrons are treated kinetically and electron temperature anisotropy develops as we expand the simulation box. This electron temperature anisotropy can impact the evolution of mirror instability [*Ahmadi et al.*, 2016a; *Ahmadi et al.,* 2016b]. The reason we chose parallel expansion is to compare our result with *Génot et al.* [2009]. Also, in parallel expansion, decrease of $\beta$ resembles the magnetosheath plasma from the bow shock to the magnetopause and the reduction in $\beta$ can eventually force the plasma into a mirror stable state according to equation (1).

We start with isotropic electrons and slightly anisotropic protons close to the mirror instability threshold according to equation (1). The simulation parameters are $n_y = n_z = 2048$, $L_y = L_z = 32 d_p$ ($d_p$ is proton inertial length), 200 particles per cell for each species, $m_p/m_e = 25$, $v_A/c = 0.1$, $a_z = (1 + q_z t')$ with the expansion parameter $q_z = 10^{-4}$. This means that at $t = 1000 \Omega_p^{-1}$, the simulation box doubles its size where $\Omega_p$ is the proton gyrofrequency. The initial plasma parameters are $T_{p\perp}/T_{p\parallel} = 1.1$, $T_{e\perp}/T_{e\parallel} = 1$, $\beta_{p\parallel} = 13$ and $\beta_{e\parallel} = 1$. The conservation of first and second adiabatic invariants lead to temperature anisotropies and generation of proton cyclotron and mirror instabilities. Protons and electrons follow the adiabatic path until the anisotropy is large enough for the instabilities to grow. Figure 1a shows the evolution of the plasma parameters in the $(\beta_{p\parallel}, T_{p\perp}/T_{p\parallel})$ plane. In order to compare the simulation results with the Vlasov linear dispersion theory prediction, we also plot the isocontours of the maximum growth rate as a function of $\beta_{p\parallel}$ and $T_{p\perp}/T_{p\parallel}$ for the mirror and proton cyclotron instabilities in the corresponding homogeneous plasma. In figure 1a, $\beta_{p\parallel}$ decreases starting with $\beta_{p\parallel} = 13$. The red solid line shows the plasma parameters in the simulation. Initially system evolves adiabatically following the green dotted line and after a transition, the plasma follows the proton cyclotron instability path which is shown by black dashed line. The system follows the marginal stability condition $\gamma_{PCI} \sim 0.01 \Omega_p$ where $\gamma_{PCI}$ is the proton cyclotron instability growth rate. We see that plasma becomes very unstable before proton cyclotron and mirror instabilities can grow. The reason is the limitation in choosing the simulation box size in PIC simulations. Therefore, we are not resolving the maximum growth rate wavelengths for small temperature anisotropies. Also, the expansion rate can affect the growth rate of the instabilities.

We start the simulation with a small electron beta ($\beta_{e\parallel} = 1$) and it decreases as we expand the simulation box. Plasma is stable to electron whistler instability [*Gary and Wang*, 1996] when we average the plasma parameters in the entire simulation domain. This leads to high electron temperature anisotropy. Figure 1b shows the evolution of proton and electron temperature anisotropies as a function of time. The presence of the electron temperature anisotropy can enhance the mirror instability growth rate and help the mirror instability grow faster than the proton cyclotron instability. One interesting feature in the electron temperature anisotropy is the bump at $350 \Omega_p t$ in figure 1b that is associated with electron heating by proton cyclotron waves [*Sironi and Narayan,* 2015].

The proton cyclotron and mirror instabilities coexist in our simulation and they both contribute in regulating the plasma but proton cyclotron instability is stronger since plasma follows its instability threshold.

## 4 Magnetic Holes

In order to measure the dominance of peaks or holes, we use a statistical value called skewness. Skewness measures the asymmetry of a distribution of a real value variable about its mean. If skewness is positive, it means that the asymmetry of the distribution is dominated by values larger than mean and if skewness is negative, the distribution asymmetry is toward values smaller than mean value. Extending the meaning of skewness to magnetic field perturbations, positive skewness means magnetic peaks are dominant and negative skewness means the perturbations are dominated by magnetic holes. A vanishing or small value skewness means perturbations are sinusoidal-like. For a sample of $n$ values, the skewness is

$$S = \frac{\frac{1}{n}\sum_{i=1}^{n}(B_i - \bar{B})^3}{\left(\frac{1}{n}\sum_{i=1}^{n}(B_i - \bar{B})^2\right)^{3/2}}$$

where $\bar{B}$ is sample mean. We measure the skewness in $B_z$. Since proton cyclotron instability propagates parallel to the background magnetic field, it can only have perturbations perpendicular to the magnetic field to keep $\nabla \cdot B = 0$. Therefore, the parallel magnetic field fluctuations ($\delta B_z$) are only due to mirror instability. Figure 2 shows the measured skewness for magnetic field fluctuations in parallel direction (red line) and distance to mirror instability threshold (blue line) calculated according to equation (1) as a function of time. As expansion proceeds, the temperature anisotropy increases and plasma becomes unstable to mirror and proton cyclotron instabilities. Mirror mode fluctuations remains mainly sinusoidal until $\Omega_p t = 250$ since skewness is zero. After this time, mirror fluctuations grow in amplitude and start shaping as peaks. The growth of mirror fluctuations reduces the distance to threshold as we see in Figure 2. About $\Omega_p t = 580$, as plasma is approaching a marginally mirror stable region, the magnetic peaks start collapsing and magnetic holes become the dominant structures. As time goes on, the magnetic fluctuations transit to periodic structures for a short period of time. At $\Omega_p t = 1180$, magnetic holes get deeper as skewness is becoming more negative and they dominate in the marginally mirror stable region.

We also plotted the skewness as a function of distance to mirror instability threshold ($R_m$) and plasma beta ($\beta_{p\parallel}$) in Figure 3 to compare with the results of hybrid expanding box simulation by *Génot et al.* [2009]. In Figure 3a, the skewness becomes negative for $R_m < 0.75$ and in Figure 3b, skewness is negative for $\beta_{p\parallel} < 2.7$. Our results are consistent with *Génot et al.* [2009] simulation and observations. In *Génot et al.* [2009] simulation, the peak to hole transition starts at $R_m \sim 0.7$ and $\beta_{p\parallel} \sim 3.1$ (see Figure 9 in their paper).

To show that the magnetic holes get very deep as skewness becomes more negative, we have made cuts through the parallel magnetic field fluctuations in Figure 4. Figures 4a and 4b show the $B_z$ fluctuations (associated with mirror instability only) at $\Omega_p t = 866$ and $\Omega_p t = 1407,$ respectively. Figures 4c to 4f show the cuts through magnetic field perturbations along y axis (blue lines) and along z axis (red lines). It is clear from these figures that the mirror mode magnetic hole is getting deeper while the magnetic peaks amplitudes doesn't change and it is saturated. If we calculate the distance to mirror instability threshold locally in the simulation box, at the late stages of the simulation, the plasma is mirror stable in magnetic peaks but it is strongly mirror unstable in magnetic holes. On average, plasma is marginally mirror stable.

Our expanding box simulation shows that magnetic peaks are dominant when plasma parameters are far from mirror instability threshold and mirror fluctuations evolve to deep magnetic holes when plasma is marginally mirror stable. The survival of the magnetic holes in a marginally mirror stable plasma agrees with bi-stability theory proposed by *Kuznetsov et al*. [2007] and *Califano et al.* [2008].

## 5 Summary and Conclusions

In this paper, we used an expanding box PIC method to study the evolution of mirror instability structures in an expanding plasma. The expansion resembles the magnetosheath plasma under dynamic solar wind pressure. This method let the proton temperature anisotropy be driven self consistently and led to the generation of mirror and proton cyclotron instabilities. We investigated the shape of mirror structures and their relations to plasma parameters in order to explain the observed mirror mode magnetic holes in the magnetosheath. Our simulation show that the shape of mirror mode structures depends on the distance to mirror instability threshold. Theoretical studies have shown that in a mirror unstable plasma, both magnetic peaks and magnetic holes can be present but only magnetic holes can survive in a mirror stable plasma. Our results are consistent with the theoretical works and also hybrid expanding box simulations. We showed that as plasma approaches the marginally mirror stable states, the mirror structures transit from magnetic peaks to magnetic holes. Magnetic peaks get saturated while magnetic holes continue to grow and get deeper. Plasma is mirror stable in magnetic peaks but it is strongly mirror unstable in magnetic holes.

For the future investigations on the mirror instability, we can study the effects of the expansion speed on the evolution of the mirror structures. A fast expansion quickly creates an anisotropic plasma and instabilities have to grow faster to overtake the expansion and regulate the plasma. This can impact the evolution of the magnetic structures. One interesting feature that we are observing in our expanding box simulations, is the electron heating by proton cyclotron waves which was reported by *Sironi and Narayan* [2015] in compressing box simulations. This is an interesting subject that can be studied further. We also observe the electron whistler signatures at the gradients of magnetic holes which agrees with recent Magnetospheric Multiscale mission observations. This topic is beyond scope of this paper and it is under preparation for future publication.

**Acknowledgments, Samples, and Data**

This work was supported by National Science Foundation grant AGS-1056898 and Department of Energy grant DESC0006670. Computations were performed using the following resources: Trillian, a Cray XE6m-200 supercomputer at UNH supported by NSF MRI program under grant

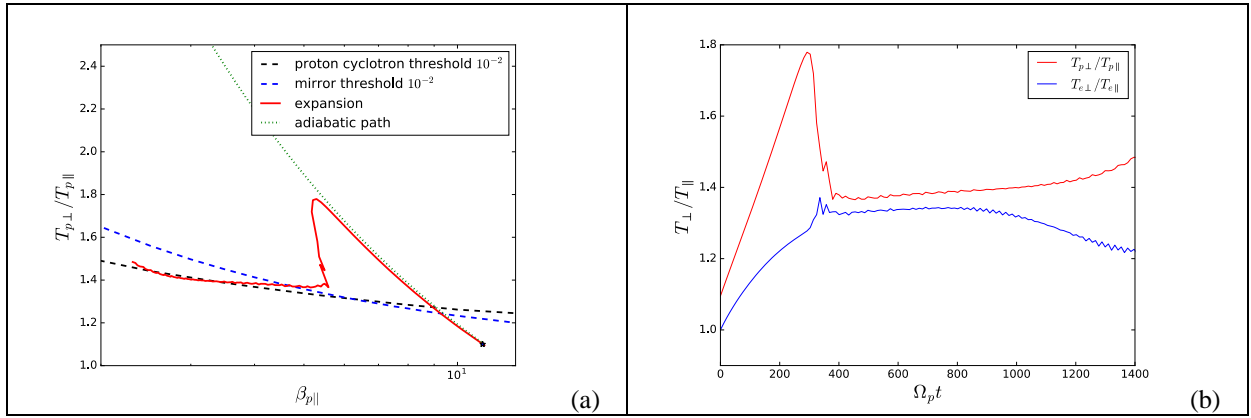

Figure 1: (a) Evolution during plasma expansion (red line) in the ($\beta_{p||}, T_{p\perp}/T_{p||}$) space. The overplotted curves show the contours of the maximum growth rate in the corresponding bi-Maxwellian plasma for mirror and proton cyclotron instabilities. (b) Evolution of electron temperature anisotropy (blue line) and proton temperature anisotropy (red line) in expanding box simulation. Both electron and proton temperature anisotropies increase following adiabatic path until temperature anisotropy instabilities start growing. There is a bump in electron temperature anisotropy at $350\Omega_p t$ that is associated with electron heating by proton cyclotron waves.

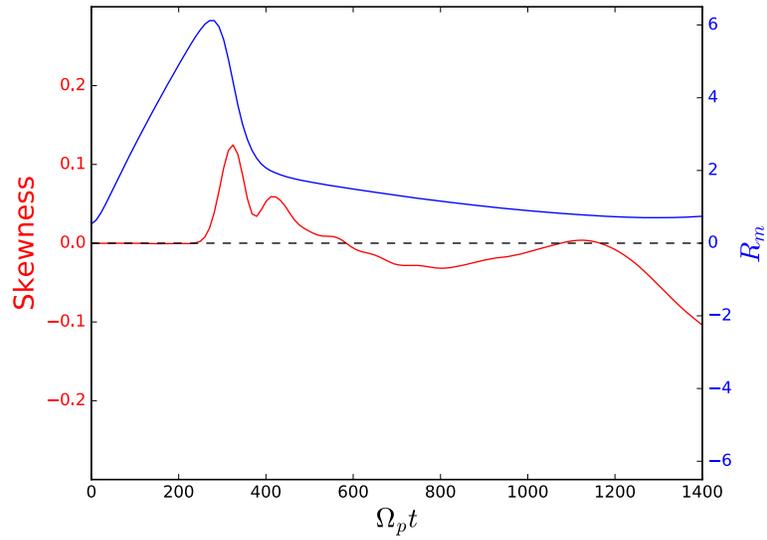

Figure 2: Skewness of $B_z$ and distance to mirror instability threshold in expanding box simulation. At early times, magnetic peaks are dominant but when plasma approaches the marginal stability threshold, magnetic holes become dominant.

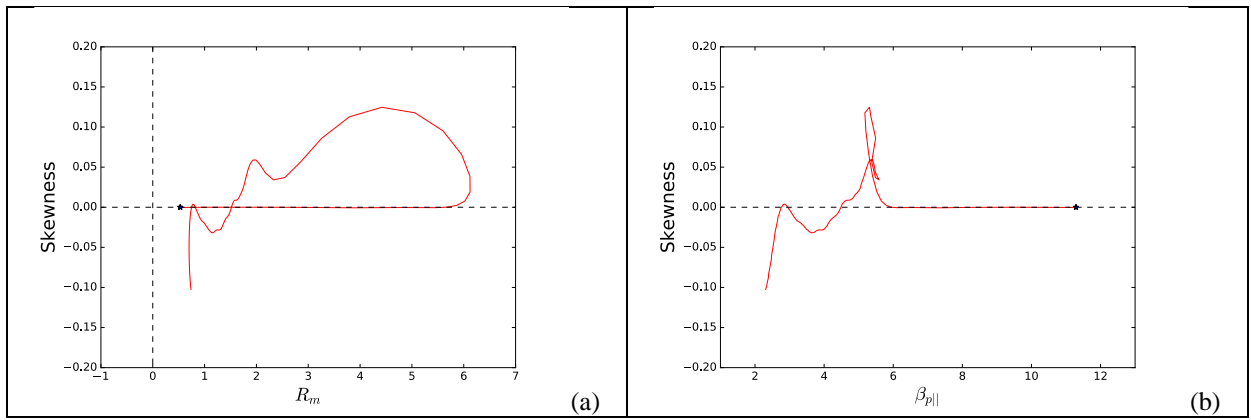

Figure 3: (a) Skewness of $B_z$ versus distance to mirror instability threshold in expanding box simulation. At early times, magnetic peaks are dominant but when plasma approaches the marginal stability threshold, magnetic holes become dominant. (b) Skewness of $B_z$ versus plasma beta in expanding box simulation. The mirror structures transit from magnetic peaks to deep magnetic holes around $\beta_{p\parallel} \sim 2.7$.

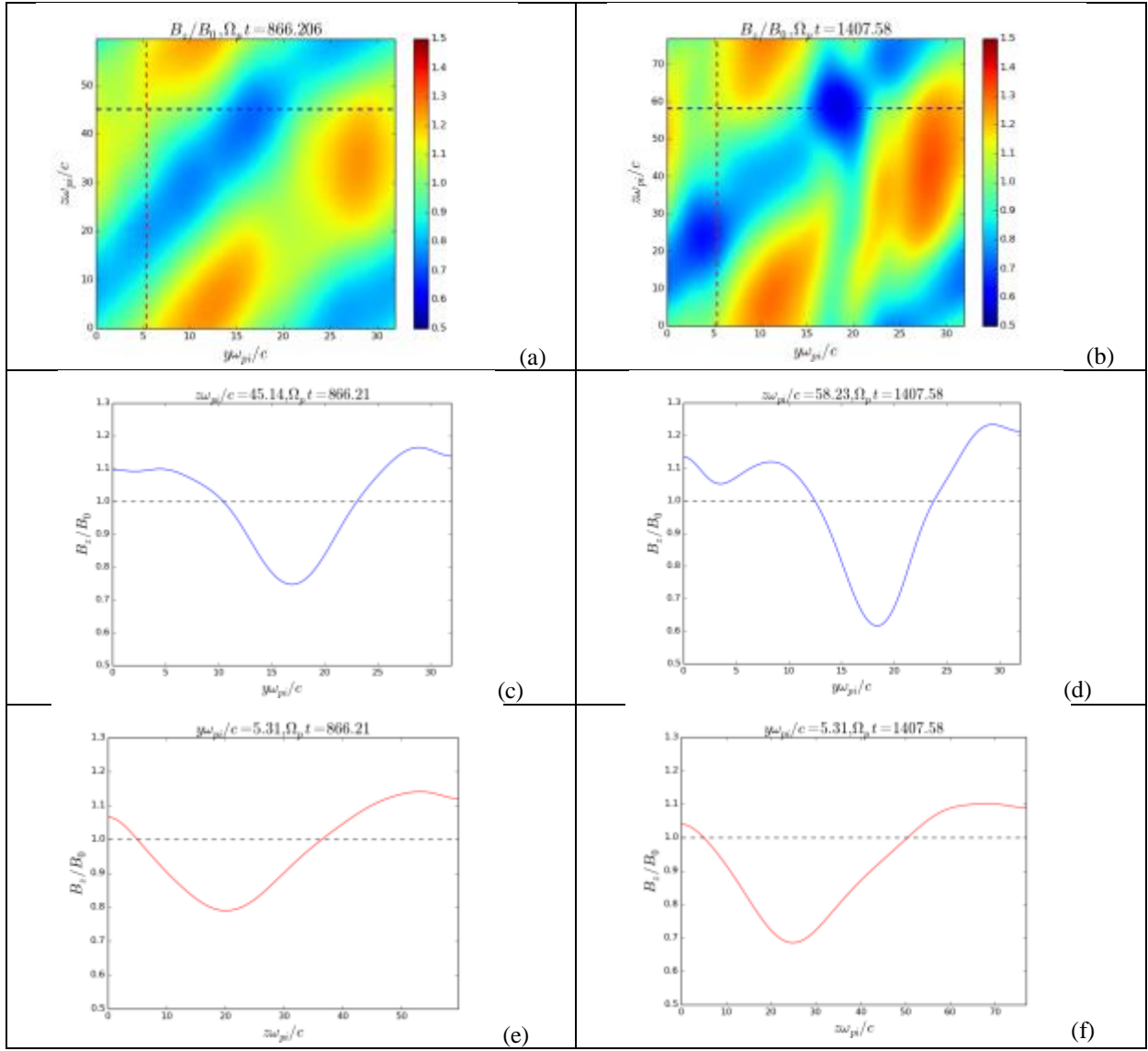

Figure 4: (a) and (b) $B_z$ fluctuations at $\Omega_p t = 866$ and $\Omega_p t = 1407$, respectively. The mirror mode fluctuations are stationary and they evolve to deep magnetic holes. (c) and (d) show a cut through $B_z$ fluctuations along y axis that is shown by dashed blue lines in first row. (e) and (f) are cuts along z axis shown by red dashed lines in first row.